\documentstyle[12pt,blois,psfig]{article}
\begin{document}
\heading{A UNIFORM PRIMORDIAL DEUTERIUM\\ 
         ABUNDANCE STEMMED FROM\\
          QSO OBSERVATIONS}

\author{Sergei A. Levshakov $^{ }$ $^{ }$} {$^{ }$ National Astronomical Observatory, 
       Mitaka, Tokyo 181-8558, Japan.} {$^{ }$ }

\begin{bloisabstract}
The discordance between `low' and `high' primordial D/H measurements can be considerably
reduced if the analysis of the H+D profiles accounts for the correlated velocity field
of bulk motion. We have re-estimated the value of D/H at $z = 3.572$ towards Q1937--1009
(`low' D/H = $(3.3 \pm 0.3)\times10^{-5}$ \cite{BT}) and at $z = 0.701$ towards Q1718+4807
(`high' D/H = $(2.0 \pm 0.5)\times10^{-4}$ \cite{JW}) and found that a single D/H value 
from the range $(3.5 - 5.2)\times10^{-5}$ $(2\sigma$ C.L.) is sufficient to describe both
spectra. This result supports homogeneous models of BBN.
The obtained D/H ratio and the measurements of
high $^4$He abundance in extra-galactic H~II regions \cite{IZ} and $^7$Li abundance in metal-poor halo
stars \cite{BO} are consistent, within the errors, 
with the predictions of standard BBN for
$3.6 \leq \eta_{10} \leq 5.4$, which corresponds to
$0.013 \leq \Omega_{\rm b}h^2_{100} \leq 0.020$.
\end{bloisabstract}

\section{Introduction}

The light elements D, $^3$He, $^4$He and $^7$Li are produced in big bang nucleosynthesis (BBN).
In the standard hot homogeneous BBN model their relative abundances depend on a single
parameter $\eta$ -- the ratio between the number of baryons and photons at the epoch of
nucleosynthesis (see e.g. \cite{HA}). The value of $\eta$ is not, however, predicted by
the big bang theory. It can only be estimated from observations. 
To fix $\eta$, the measurements of the (D/H)$_p$ ratio (the subscript $p$ throughout
denotes primordial abundance) can be carried out from spectra of distant quasars \cite{AD}.
Such measurements, together with other three light element abundances, 
provide the complete test of the standard BBN model.

Recent high-redshift D abundances are, however, very discrepant (cf. \cite{BT} and \cite{JW}).
It was suggested that there may be inhomogeneity in (D/H)$_p$ of at least a factor of ten.
This implies that big bang nucleosynthesis may have occurred inhomogeneously \cite{JW} 
and there may not be a unique primordial deuterium abundance as first pointed out in \cite{JE}.

In these proceedings, we show one possibility how to reconcile the observed (D/H)$_p$
values. We demonstrate that the discrepancy between `high' and `low' D/H measurements
may be overcome in the framework of the line broadening process
which accounts for the spatial correlations in the intervening velocity field.

\section{The D/H measurements}

We consider two examples of real data which yielded two limiting D/H values 
in the previous studies.
The first one is the absorbing region with observable hydrogen and deuterium Ly$\alpha$ and Ly$\beta$
lines at $z = 3.572$ towards Q1937--1009 where a `low' D/H ratio of
$(3.3 \pm 0.3)\times10^{-5}$ was measured in \cite{BT}.
The second one is the H+D Ly$\alpha$ blend with D/H = $(2.0 \pm 0.5)\times10^{-4}$ 
seen at $z = 0.701$ towards Q1718+4807 \cite{JW}. The latter D/H value
is at the upper end of the D/H range measured from QSO spectra. 

The foregoing estimations were obtained through the Voigt profile fitting analysis.
This analysis ignores the large-scale correlations in the velocity field and thus
may produce very misleading results when applied to the case where the spatial
correlations in bulk motion are significant (see \cite{LT1}, \cite{LT2}, and \cite{LKM}).

Our method takes into account the correlated structure of the radial velocity field.
The model supposes a continuous absorbing region of a thickness $L$ exhibiting a mixture 
of bulk motions of different types. 
The gas motion along a given line of sight is described
by a fluctuating (random) velocity field. For the sake of simplicity,
we assume a homogeneous (H~I) density and kinetic temperature T$_{kin}$.
The velocity field is characterized by its rms amplitude  $\sigma_{t}$
and a correlation length $l > 0$
(the reader is referred to \cite{LKT} for more details).

To estimate physical parameters and an appropriate velocity field
structure along the sight-line, $v(s)$, we used a reverse Monte Carlo (RMC)
technique. Our algorithm requires the definition of
a simulation box for the five physical parameters~:
N(H~I), D/H, T$_{kin}$, $\sigma_{t}/v_{th}$, and $L/l$
(here $v_{th}$ denotes the thermal width of the hydrogen lines).
The velocity, which is the continuous random function of
the coordinate $s$, is represented by
its sampled values at equally spaced intervals $\Delta s$, i.e. by the vector
$\{v_1, v_2, \dots , v_k\}$
of the components parallel to the line of sight
at the spatial points $s_j$ (see \cite{LKT}).
 
The obtained results for the $z = 3.572$ and $z = 0.701$ systems are summarized in Figure~1. 
Upper left panel in this figure shows confidence regions 
in the `N(H~I)--D/H' plane for the simultaneously 
fitted blue wings of the H+D Ly$\alpha$ and Ly$\beta$ profiles (Q1937-1009)
when the other parameters 
(T$_{kin} = 1.15\times10^4$ K, $\sigma_t/v_{th} = 1.46$, and $L/l$ = 3.6) and 
the corresponding configuration of the velocity field $v(s)$ are fixed. 
The contours represent 68.3\% (innermost), 95.4\%, and 99.7\% (outermost)
confidence levels. The filled circle 
marks the point of maximum likelihood for model (d) from \cite{L1}
where $\chi^2_{min}$ per degree of freedom is 1.18. 
Lower left panel demonstrates the confidence range for D/H from the
second system (Q1718+4807). The vertical dashed lines restrict the N(H~I)$_{total}$ value
in accord with \cite{JW}. The maximum likelihood point (filled circle) corresponds to
D/H~= $5.4\times10^{-5}$, N(H~I)~= 1.71$\times10^{17}$ cm$^{-2}$, T$_{kin}~= 1.5\times10^4$~K,
$\sigma_t~= 26$ km~s$^{-1}$, and $L/l~= 3.5$ with $\chi^2_{min}~= 0.804$. This point
is slightly shifted with respect to the best model (e) from \cite{L2} with $\chi^2_{min} = 0.897$
(open circle)
because in the present study a wider simulation box for the physical parameters was adopted.
The contours in this panel represent 68.3\% (inner) and 95.4\% confidence levels.

These two measurements of D/H in the high-redshift systems reveal no variations of the
deuterium abundance within the quality of the data sets. We can, therefore, compare
SBBN predictions with observational abundances
which are shown in Figure~1 as well. 
The theoretical SBBN relative abundances of D, $^4$He, and $^7$Li
(solid curves in the right panels) and their uncertainties (dashed curves) as
function of $\eta$ are taken from \cite{SA}.
The abundances of D and $^7$Li are number ratios, whereas $Y_p$ is the mass fraction of
$^4$He.  In the upper right panel, the rectangle height 
corresponds to the $2\sigma$ uncertainty region for D/H from the $z = 3.572$ system
(which provides the highest quality data to date). 
We do not show here the corresponding bounds from the $z = 0.701$ system since its
H+D profile was obtained with seven times lower signal-to-noise ratio resulting
in a wider D/H uncertainty range which entirely overlaps the $z = 3.572$ 
confidence interval for D/H.
The rectangle heights in
two lower panels give the bounds from recent measurements of extra-galactic $Y_p$
in metal-poor H~II regions
\cite{IZ} and $^7$Li in warm metal-poor population~II (halo) stars \cite{BO}. 
We do not depict an error box representing another $^4$He abundance measurement \cite{OL},
since ($i$) our D/H value is not consistent with $Y_p = 0.234 \pm 0.002$ found in \cite{OL} and
($ii$) the estimations of $Y_p$ in \cite{IZ} and \cite{OL} are not in agreement.
The rectangle widths give the 
$2\sigma$ uncertainty regions for $\eta$ in the SBBN calculations. The shaded region is the
window for $\eta$ common to all results. It specifies $2\sigma$ confidence levels 
in $\eta$. 

Taken at face values, the shaded $\eta$-region $3.6 \leq \eta_{10} \leq 5.4$
leads directly to the estimation of the
present-day baryonic density. Using the present-day number density of photons of the
cosmic background radiation (T~= 2.73~K), we can obtain
$0.013 \leq \Omega_{\rm b}h^2 \leq 0.020$, where
$H_0 = 100h$~km~s$^{-1}$~Mpc$^{-1}$.

Now we can conclude that in the framework of our model we do not detect
any variations of the D/H ratio. This result supports SBBN. However,
more information on the H+D absorption-line systems is needed to verify
this statement unambiguously.

\acknowledgements{I thank W. Kegel and F. Takahara for their cooperation,
D. Tytler and J. Webb for providing the spectra of Q1937--1009 and Q1718+4807,
respectively.
I am also grateful to the Minist\`ere de l'Education Nationale, de la Recherche
et de la Technologie (MENRT) for funding my participation to the conference.
This work was supported in part by the Grant-in-Aid of Foundation for Promotion 
of Astronomy, Japan.}

\vfill

\newpage

\begin{figure*}
\vspace{0.5cm}
\hspace{0.0cm}\psfig{figure=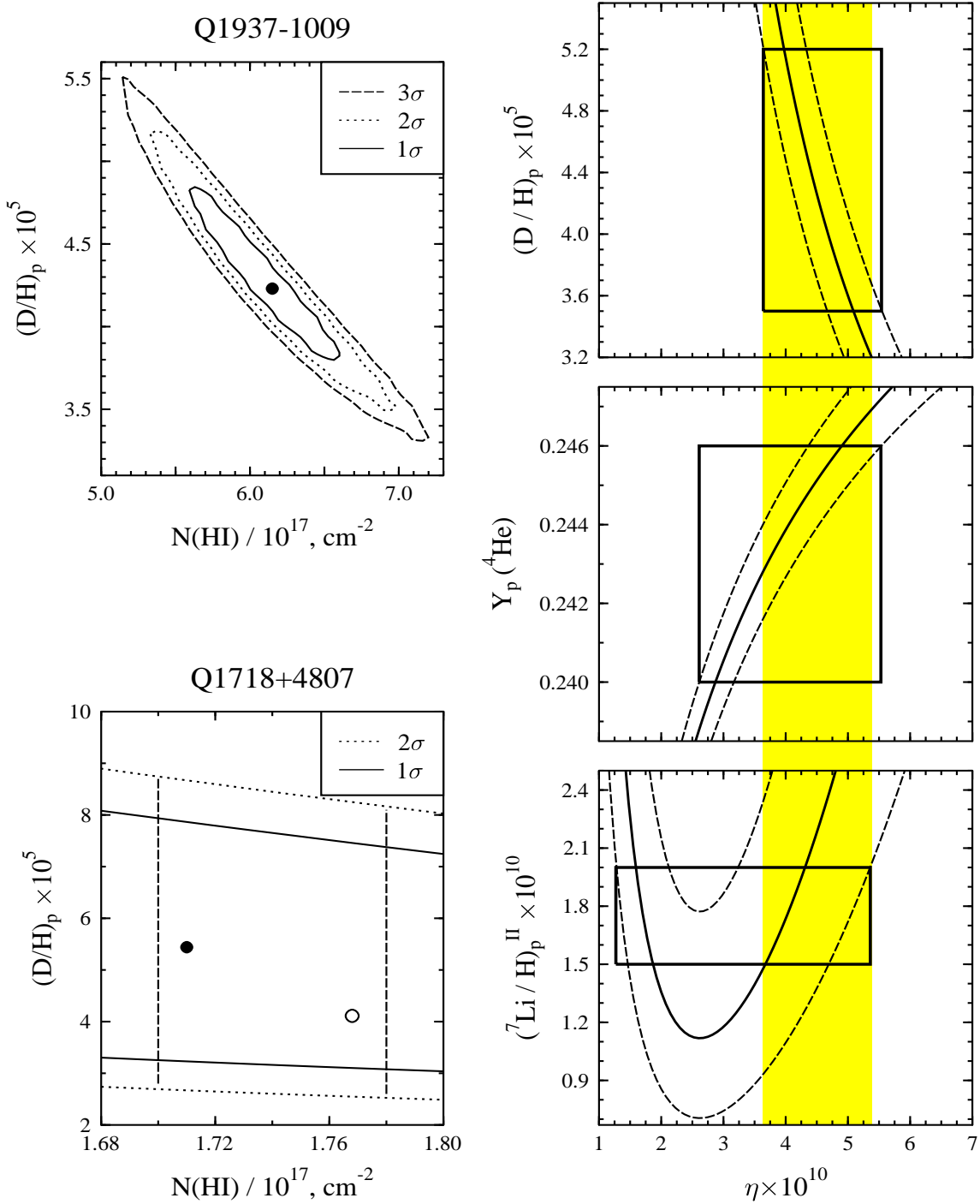,height=23.0cm,width=16.5cm}
\vspace{-3.0cm}
\caption[]{ {\it Upper left}~-- Confidence regions for the 
simultaneously fitted H+D Ly$\alpha$ and Ly$\beta$
profiles from the $z = 3.572$ system towards Q1937--1009. {\it Lower left}~-- 
Confidence regions for the fitted H+D Ly$\alpha$ profile observed at $z = 0.701$ towards
Q1718+4807. {\it Right}~-- Comparison of predicted by SBBN primordial abundances with observational
bounds. See text for more details. }
\end{figure*}

\end{document}